\documentclass[aps,prb,twocolumn,superscriptaddress]{revtex4-2}

\usepackage{graphicx}
\usepackage{dcolumn}
\usepackage{bm}
\usepackage{mathptmx}
\usepackage{array} 
\usepackage[version=4]{mhchem}
\usepackage[colorlinks=true, letterpaper=true, linkcolor=blue, citecolor=blue, urlcolor=blue]{hyperref}
\usepackage{multirow}
\usepackage[mathlines]{lineno}

\begin{document}


\title{Integration of promising piezoelectric and photocatalytic properties in Janus In$XY$ ($X$ = S, Se, Te; $Y$ = Cl, Br, I) monolayers and their heterojunctions}

\author{Xinyue Liu}
\author{Ziqiang Li}
\author{Yanfeng Ge}
\author{Yong Liu}
\affiliation{State Key Laboratory of Metastable Materials Science and Technology $\&$ Hebei Key Laboratory of Microstructural Material Physics, School of Science, Yanshan University, Qinhuangdao 066004, P. R. China.}
\author{Xing Wang}
\affiliation{College of Science, Hebei North University, Zhangjiakou 07500, P. R. China}

\author{Wenhui Wan}%
\email[Corresponding author: ]{wwh@ysu.edu.cn}
\affiliation{State Key Laboratory of Metastable Materials Science and Technology $\&$ Hebei Key Laboratory of Microstructural Material Physics, School of Science, Yanshan University, Qinhuangdao 066004, P. R. China.}




\date{\today}

\begin{abstract}
Two-dimensional (2D) Janus materials show great promise as piezoelectric materials and photocatalysts for water splitting. In this work, we systematically investigated the piezoelectric and photocatalytic properties of the hexagonal Janus In$XY$ ($X$ = S, Se, Te; $Y$ = Cl, Br, I) monolayers (MLs) using first-principles calculations. Except for InSeCl ML, the remaining eight In$XY$ MLs are stable and exhibit exceptionally high in-plane piezoelectric coefficients ($|d_{22}|$ = 6.07--155.27 pm/V), 
which exceed those of most known 2D materials. In$XY$ MLs possess band edges straddling the water redox potentials at pH = 0. Their intrinsic vertical polarization induces an intralayer polarization field $E_{\rm intra}$, leading to low exciton binding energies (0.44--0.78 eV). Moreover, their strong vertical piezoelectric responses ($|d_{32}|$ = 0.34--0.65 pm/V) suggest that in-plane stress can further enhance $E_{\rm intra}$ to facilitate the separation of photogenerated carriers. Additionally, these In$XY$ MLs exhibit high electron mobility (101--899 cm$^2$/V/s) and a pronounced anisotropy ratio in carrier mobility, which effectively suppresses charge recombination. 
Among them, several stand out: InSI and InSeBr MLs show high electron mobility and a large carrier mobility anisotropy ratio; InSeBr ML exhibits excellent in-plane and out-of-plane piezoelectricity; and InSeBr, InSeI, and InTe$Y$ ($Y$ = Cl, Br, I) MLs show strong visible-light absorption. To optimize performance, we constructed a van der Waals heterojunction (InSI/InSeBr), which demonstrates remarkable photocatalytic properties, including enhanced redox ability, a direct Z-scheme charge transfer pathway, strong visible-light absorption, high carrier mobility, and excellent photocorrosion resistance. Our results indicate that hexagonal Janus In$XY$ MLs and their heterojunctions are multifunctional materials integrating piezoelectricity and photocatalysis, paving the way for energy conversion applications.
\end{abstract}


\maketitle


\section{INTRODUCTION}
The rapidly growing global economy accelerates the extraction and consumption of natural resources, which are significant causes of the triple planetary crisis: climate change, biodiversity loss, and pollution. The technology of photocatalytic water splitting~\cite{Nishioka2023} generates clean hydrogen (H$_2$) from abundant sources: water and solar light, promoting a green transformation of resource utilization mode~\cite{AHMAD2023102830,LI2022111980}. Fujishima et al. (1972) demonstrated that semiconducting TiO$_2$ can absorb light to generate charge carriers for water splitting and hydrogen production~\cite{fujishima1972electrochemical}. However, conventional semiconductor photocatalysts suffer from rapid electron-hole pairs recombination, low utilization of light energy, and existing photocorrosion phenomenon~\cite{takata2020photocatalytic,ge2017review}. The development of efficient photocatalysts is urgent. 
Recent studies have revealed that photocatalysts with piezoelectricity, named piezo-photocatalysts, can achieve efficient hydrogen production~\cite{Jia2024,Yu2023,jing2023piezo}. In these materials, light excitation generates electron-hole pairs, while mechanical stress induces or enhances a polarization-driven built-in electric field (polarization field) to promote the separation and migration of photo-generated carriers~\cite{Tu2020,Masekela2025}. Such synergistic effects make piezo-photocatalysts promising candidates for practical water splitting devices, with polarization emerging as a key performance determinant \cite{Zhu2021a}. However, finding and designing high-performance piezo-photocatalysts still face challenges.

The prevalent piezo-photocatalysts adopt the form of core-shell heterojunctions, with piezoelectric materials constituting the core and visible-light photocatalysts forming the shell~\cite{2024ResearchPO}. Heterojunction can combine the intrinsic features of the component materials to synergistically enhance the photocatalytic activity~\cite{low2017heterojunction}. Hong et al. created a ZnO/CuS heterojunction in which the polarization field within ZnO and the interfacial electric field separate photogenerated carriers, thus significantly suppressing recombination rates~\cite{hong2016high}. Liu et al. constructed a \ce{BaTiO3 / ReS2} heterojunction in which \ce{ReS2} enhances light absorption and the polarization field of \ce{BaTiO3} reduces the Schottky barrier for transferring photogenerated electrons to \ce{ReS2}~\cite{liu2022directing}. 

Since the synthesis of graphene and other two-dimensional (2D)  materials~\cite{ARES20223,novoselov2005two}, van der Waals (vdW) heterojunctions have been fabricated by stacking two or more different 2D layered materials with weak interlayer coupling. Recently, vdW heterojunctions with the type-II band alignment and Z-scheme carrier transfer pathway have been proposed for photocatalysis~\cite{Hu2024}. vdW heterojunctions containing piezoelectric materials possess intralayer polarization fields and interfacial electric fields. The coupling of these two types of electric fields can modulate and promote photocatalytic performance. Meanwhile, the vertical assembly of vdW heterojunction preserves the catalytic sites of stacked nanolayers~\cite{xu2017enhanced,guan2022boosting}, improves the optical absorption ability~\cite{wang20222d,yan2020van}, adjusts the electronic structure~\cite{Zhu2021,Li2023,yang2021short}, and improves carrier separation and transport~\cite{Chen2022}. However, identifying suitable 2D piezoelectric materials is an essential prerequisite for constructing vdW heterojunctions as efficient piezo-photocatalysts.

In 2017, Lu et al.~\cite{lu2017janus} and Zhang et al.~\cite{zhang2017janus} successfully synthesized Janus MoSSe monolayer (ML) using a modified CVD method~\cite{lu2017janus,zhang2017janus}. 2D Janus materials possess both in-plane and out-of-plane piezoelectricity~\cite{guo2017enhanced,dong2017large}. 
Many Janus MLs exhibit high piezoelectric coefficients~\cite{ng2018graphene, guo2017enhanced}, showing potential applications in photocatalysis~\cite{Wu2018}. Ju et al. predicted that Janus WSSe ML possesses excellent photocatalytic properties, including appropriate band edge positions for water-splitting, high visible-light absorbance, a small exciton binding energy of 0.83 eV, a solar-to-hydrogen (STH) efficiency of 11.68\%, and good resistance against the photo-induced corrosion~\cite{Ju2020}. Yang et al. predicted that Janus \ce{ZrOBrI} ML has a high transverse piezoelectric coefficient of $d_{21} = 2.53$ pm/V, suitable band edge positions, orientation-dependent optical absorption, and anisotropic carrier transport~\cite{Yang2024}.  Zhao et al. proposed a series of Janus semiconductors B$XY$ ($X$ = P, As, Sb; $Y$ = S, Se, Te) with superior carrier mobility, optical absorption properties, and STH efficiency~\cite{Zhao2025}. 2D Janus materials can stimulate the advancements in piezo-photocatalysis technology. Nonetheless, the application of 2D Janus materials as photocatalysts remains constrained. Beyond generating a substantial built-in polarization field for effective charge separation, these materials must simultaneously satisfy multiple criteria: appropriate band edge positions for redox potentials, strong visible-light absorption, and fast carrier delivery to catalytic sites for redox reactions. Consequently, 
ongoing efforts are needed to discover novel 2D Janus materials as potential piezo-photocatalysts for water splitting.

Recently, 2D In-monochalcogenides In$X$ ($X$ = O, S, Se, Te), whose hexagonal lattice consists of a quadruple layer in a stacking sequence of $X$–In–In–$X$, have been synthesized~\cite{KakanakovaGeorgieva2021,Tu2020a,Bandurin2017,Matetskiy2022}. These materials exhibit promising electronic properties such as layer-dependent band gaps, high electron mobility, and suitable band edge positions for photocatalytic water splitting \cite{Cui2018}. However, their symmetric lattices lack intrinsic polarization, which hinders the efficient separation of photogenerated charge carriers. 
One strategy to introduce vertical polarization into these monolayers is the formation of Janus structures by substituting In atom or chalcogen $X$ atom with isoelectronic elements \cite{Liu2021}. 
Our previous work \cite{Wan2019} shows that the conduction band minimum of 2D In$X$ is dominated by the $s$ orbital of In atoms. To maintain the excellent electron transport properties of 2D In$X$, we proposed an alternative method to obtain Janus structures from 2D In$X$ via surface halogenation.
Halogenation (i.e., fluorination, chlorination, bromination, and iodination)
has become a versatile strategy for chemical modifications of 2D materials \cite{Tang2021,Li2021,Kamysbayev2020,Liu2024}.
Structural relaxation calculations show that the halogenation can spontaneously decompose the hexagonal crystal lattice of In$X$ ($X$ = O, S, Se, Te) MLs, leading to the formation of twelve distinct Janus In$XY$ ($X$ = O, S, Se, Te; $Y$ = Cl, Br, I) MLs (see Fig. S1~\cite{supplementary} of the Supplemental Material(SM)). This process is exothermic (see Table S1~\cite{supplementary}), suggesting strong experimental feasibility.
Besides, we noticed that the bulk hexagonal phase of InSCl, InSBr, InSeCl, InSeBr are exist naturally~\cite{Hahn1962}, and few-layer InSeBr has been successfully exfoliated from bulk crystals~\cite{hu2020raman}, further supporting the realizability of these Janus structures.
Meanwhile, in prior research~\cite{guo2023}, we have predicted a series of Janus Ga$XY$ ($X$ = S, Se; $Y$ = Cl, Br, I) MLs with strong piezoelectric coefficients and appropriate band edge positions for photocatalytic water splitting, hinting at similar potential in Janus In$XY$ systems. Moreover, these In$XY$ MLs can be assembled into vdW heterojunctions by combining two distinct Janus layers. The relevant works on photocatalytic properties of vdW heterojunctions with two Janus MLs are still limited \cite{Ju2022,Liu2021,Ye2023}.
The piezoelectric and photocatalytic properties of InSeBr-type Janus structures and related vdW heterojunctions have not been thoroughly investigated, making it crucial to fill this research gap.

Inspired by that, in this work, we systematically investigated the piezoelectric and photocatalytic properties of In$XY$ ($X$ = O, S, Se, Te; $Y$ = Cl, Br, I) MLs by first-principles calculations. 
Our results showed that eight In$XY$ MLs (excluding InO$Y$ ($Y$ = Cl, Br, I) and InSeCl MLs) exhibit energetic favorability, semiconducting behavior, structural stability, and both in-plane and out-of-plane piezoelectricity.
We examined their suitability for photocatalytic applications based on band edge positions, optical absorption, electron-hole separation, and carrier mobility. All the eight In$XY$ MLs are identified as potential piezo-photocatalysts for overall water splitting. To further improve the photocatalytic performance, we designed a InSI/InSeBr heterojunction with the I-Se interface. This heterojunction demonstrates synergistic advantages: enhanced piezoelectric responses, a Z-scheme charge transfer pathway for efficient carrier separation, an optimal 2.330 eV band gap, high charge mobility, enhanced visible-light absorption, and good photocorrosion resistance. These results reveal that In$XY$ MLs and their vdW heterojunctions can be utilized as piezo-photocatalysts for water splitting.

\section{COMPUTATIONAL METHODS}
First-principles calculations were performed by the Vienna ab initio simulation package (VASP)~\cite{q28} with the projector augmented wave (PAW)~\cite{q29} pseudopotentials and Perdew, Burke, and Ernzerhof (PBE)~\cite{q30} exchange-correlation functional. A vacuum layer of 20 \AA\ was employed to prevent artificial interaction between the adjacent periodic images. During the calculations of the vdW heterojunctions, the DFT-D3 method by Grimme~\cite{grimme2010consistent} was used to correct the vdW interactions. The kinetic energy cutoff is 550 eV. The Brillouin zone was sampled using a 12 × 12 × 1 uniform $\bf k$-point mesh generated via the Monkhorst-Pack scheme~\cite{q31}. The convergence criteria of total energy and atomic force were set to 10$^{-6}$ eV and 10$^{-3}$ eV/\AA, respectively. Band structures were calculated using the Heyd-Scuseria-Ernzerhof (HSE06) hybrid functional~\cite{q32}. Electronic polarization was calculated by the Berry phase method~\cite{q34}. Phonon dispersion was computed with a $4\times4\times1$ supercell using the finite displacement method, as implemented in the Phonopy code~\cite{q33}. Molecular dynamics (MD) simulations were performed with a $4\times4\times1$ supercell in the canonical (NVT) ensemble at 300 K for 7 ps with a time step of 2 fs.

According to Voigt's notation~\cite{Nye1957}, the six components of strain matrix $\varepsilon$ including $\varepsilon_{xx}$, $\varepsilon_{yy}$, $\varepsilon_{zz}$, $\varepsilon_{yz}$($\varepsilon_{zy}$), $\varepsilon_{xz}$($\varepsilon_{zx}$), and $\varepsilon_{xy}$($\varepsilon_{yx}$) can be abbreviated as $\varepsilon_{j}$, $j$=1--6. 
For a 2D hexagonal lattice, the independent planar relax-ion elastic constants $C_{11}$ and $C_{12}$ were determined by fitting the energy $U$ of the unit cell to in-plane strains ($\varepsilon_{1}$, $\varepsilon_{2}$):
\begin{equation}
	C_{11}= \frac{1} {A_{0}} \frac{\partial^2U} {\partial\varepsilon^2_{1}},
	C_{12}= \frac{1} {A_{0}} \frac{\partial^2U} {\partial\varepsilon_{1}\partial\varepsilon_{2}}.
	\label{stiffness}
\end{equation}
Here, $A_{0}$ is the area of the unit cell at zero strain. We took the strains $\varepsilon_{1}$ and $\varepsilon_{2}$ along the $x$- and $y$-directions from -0.01 to 0.01, with a step of 0.005.

The piezoelectric stress tensor $e_{ij}$ and piezoelectric strain tensor $d_{ij}$ are defined as
\begin{eqnarray}
	e_{ij} &=& \frac{\partial P_{i} } {\partial \varepsilon_{j}}=  \sum_{k} \frac{\partial P_{i} } {\partial \sigma_{k}}\frac{\partial \sigma_{k} } {\partial \varepsilon_{j}} = \sum_{k} d_{ik}C_{kj},
    \label{eij}
	\\
	d_{ij} &=& \frac{\partial P_{i} } {\partial \sigma_{j}}.
	\label{strain}
\end{eqnarray}
Here $P_{i}$ is the $i$-th component of 2D polarization vector (dipole moment per unit area).
$\varepsilon_{i}$ and $\sigma_{i}$ are the strain tensor and the stress tensor, respectively.

The exciton binding energy $E_{b}$ was calculated by~\cite{Fu2018}
\begin{equation}
	E_{b}= E_g^{\rm G_0W_0} - E_{\rm opt}
	\label{eb}
\end{equation}
where $E_g^{\rm G_0W_0}$ and $E_{\rm opt}$ are the direct quasi-particle band gap and the energy of the first optical absorption peak obtained in G$_0$W$_0$+BSE method, respectively~\cite{Ramasubramaniam2012}. For the G$_0$W$_0$ calculations, a $15 \times 15 \times 1$ $\bf k$-grid was selected for sampling the Brillouin zone, while the cutoff energy and energy convergence were set to 200 eV and $10^{-8}$ eV, respectively. The eight highest valence bands and eight lowest conduction bands were used as the basis for excitonic eigenstates in the BSE calculations. 

The carrier mobility along the $x$-direction was calculated using modified deformation potential theory that considers the anisotropy in effective mass, elastic constants, and deformation potential~\cite{lang2016mobility}:
\begin{equation}
	\mu_{x} = \frac{e\hbar^{3}(\frac{5C_{11}+3C_{22}}{8})} {k_{B}T(m^{*}_{x})^{\frac{3}{2}}(m^{*}_{y})^{\frac{1}{2}}(\frac{9E^{2}_{lx}+7E_{lx}E_{ly}+4E^{2}_{ly}}{20})}.
	\label{mobility}
\end{equation}
Here, $e$ is the electron charge, $\hbar$ is the reduced Planck constant, $C_{11(22)}$ is the elastic constant in the $x(y)$-direction, $k_{B}$ is the Boltzmann constant, and $T$ is temperature. $m^{*}_{x(y)}$ is effective mass of the carrier in the $x(y)$-direction. $E_{lx(y)}$ is the carrier deformation potential constant in the $x(y)$-direction. 

\section{Results and discussion}
\subsection{Structural properties}

\begin{figure}[tbp!]
	\centerline{\includegraphics[width=0.47\textwidth]{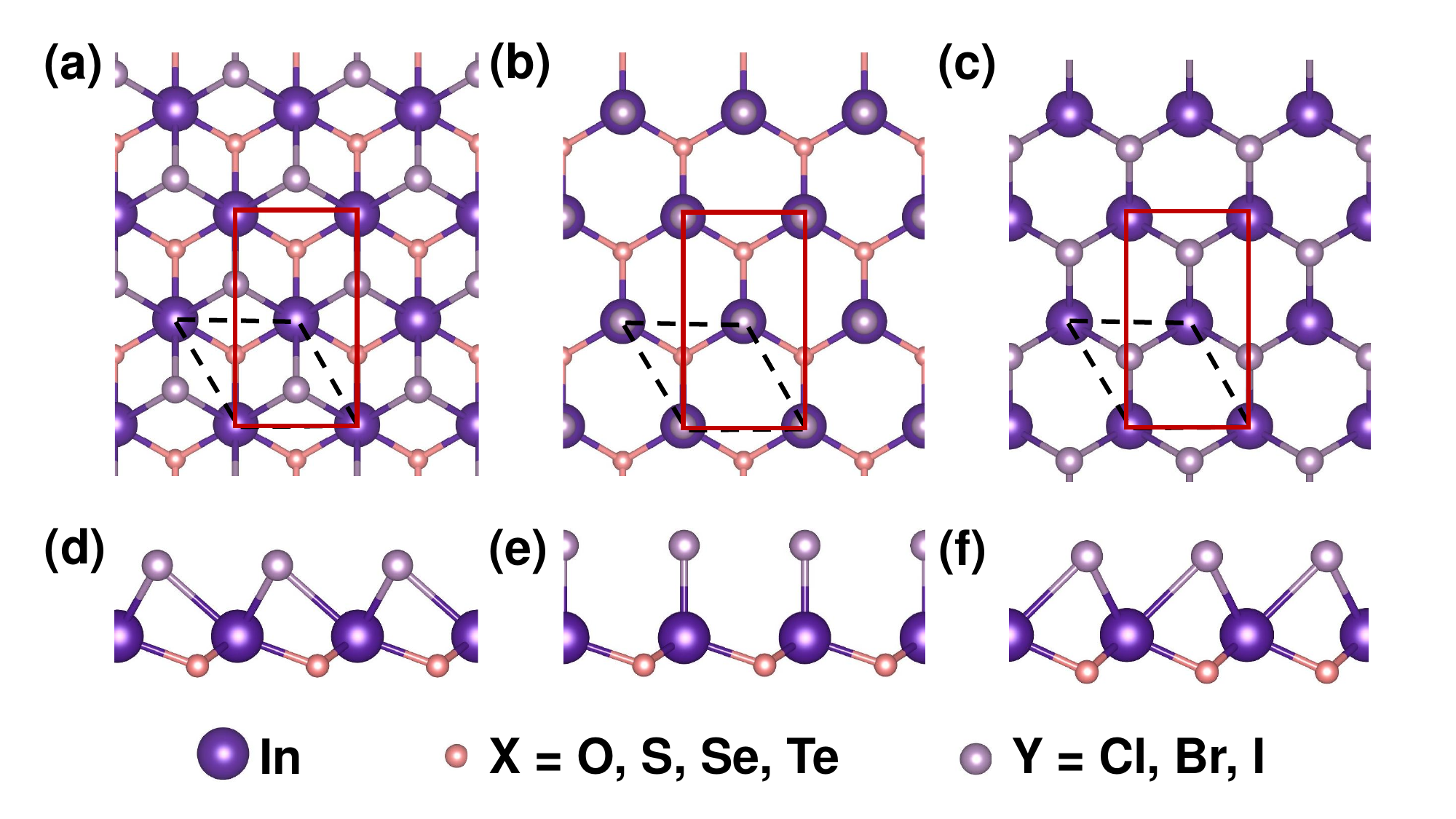}}
	\caption{Top (a)-(c) and side (d)-(f) views of In$XY$ ($X$ = O, S, Se, Te; $Y$ = Cl, Br, I) MLs in hexagonal phases I, II, and III, respectively. The primitive cell is outlined by black dashed lines and the $1 \times \sqrt{3}$ rectangular cell is marked with red solid lines.}
	\label{wh1}
\end{figure} 

\begin{table}[tbh!]
	\caption{
    Relative energies (eV per formula unit) of three hexagonal phases of In$XY$ ($X$ = O, S, Se, Te; $Y$ = Cl, Br, I) MLs. The energy of the lowest-energy phase is set to zero.}
	\label{table1}
	\centering 
\renewcommand\arraystretch{1.3}
	\begin{ruledtabular}
		\begin{tabular}{cccc}
			& Phase I   &Phase II &Phase III \\
			\hline
			InOCl &0&{0.561}&0.653\\
			InOBr &0&{0.391}&0.587\\
			InOI  &0.011&{0}&0.425\\
			InSCl &0&{0.194}&0.574\\
			InSBr &0&{0.155}&0.543\\
			InSI  &0&{0.077}&0.529\\
			InSeCl&0&{0.072}&0.549\\
			InSeBr&0&{0.064}&0.523\\
			InSeI &0&{0.021}&0.520\\
			InTeCl&0.169&{0}&0.692\\
			InTeBr&0.126&{0}&0.633\\
			InTeI &0.102&{0}&0.618\\
		\end{tabular}
	\end{ruledtabular}
\end{table}

Based on the known lattice structure of InSeBr ML \cite{shi2022ultrahigh}, we considered three possible phases for In$XY$ ($X$ = O, S, Se, Te; $Y$ = Cl, Br, I) MLs, as shown in Fig.~\ref{wh1}. 
Here, $X$ and $Y$ represent the chalcogen and halogen atoms, respectively. We excluded the In$XY$ MLs containing the Fluorine (F) atom in this study because they are unstable.
In$XY$ MLs exhibit a hexagonal lattice with a triple atomic layer structure in an $X$-In-$Y$ atomic sequence along the z-axis. 
The lattice vectors of the primitive cell are $\vec{a}_1 = a(1,0)$ and $\vec{a}_2 = a(-1/2, \sqrt{3}/2)$, where $a$ is the lattice constant.
The unequal positions of the halogen atom $Y$ distinguish the three phases. In phases I, II, and III, the halogen atom $Y$ is located at the hollow site, the top of the In atom, and the top of the $X$ atom, respectively. We relaxed the crystal lattice and compared the energy of different phases for each In$XY$ ML. 
Table~\ref{table1} shows that InSCl, InSBr, InSeCl, and InSeBr MLs adopt phase I as their lowest-energy states, which is consistent with previous studies~\cite{shi2022ultrahigh,Hu2023}. In addition, InOCl, InOBr, InSI, and InSeI MLs are also likely to form phase I. In contrast, the remaining InOI, InTeCl, InTeBr, and InTeI MLs favor Phase II as their lowest-energy states.  

\begin{table*}[bth!]
	\caption{The lattice constant ($a$), band gap ($E_{g}$) calculated by HSE06 method, gap type ($G_{T}$), relaxed-ion elastic stiffness coefficients ($C_{11}$ and $C_{12}$), Young's Modulus ($E$), 2D polarization $P_z$ (the dipole moment per area), vacuum level difference between the $X$ and $Y$ sides ($\Delta\phi$), the spatial extent of the intralayer polarization field along the $z$-axis ($h$), intralayer polarization field ($E_{\rm intra}$), exciton binding energy ($E_b$), piezoelectric coefficients ($e_{22}$, $e_{32}$, $d_{22}$, $d_{32}$) of eight In$XY$ MLs.}
	\label{table2}
	\centering 
	\renewcommand\arraystretch{1.4}
	\begin{tabular*}{1\textwidth}{@{\extracolsep{\fill}}*{16}{c}}
		\hline\hline
		&$a$     &$E_{g}$  &$G_{T}$  &$C_{11}$  &$C_{12}$  &$E$ & $P_z$ &$\Delta\phi$ &$h$ &$E_{\rm intra}$ & $E_b$ &$e_{22}$  &$e_{32}$  &$d_{22}$  &$d_{32}$ \\ \cline{5-7} \cline{13-14} \cline{15-16}
		
		& (\AA)      &(eV)     &     &\multicolumn{3}{c}{(N/m)}       & (pC/m)  & (eV) &(\AA) &(V/nm) & (eV)  &\multicolumn{2}{c}{(pC/m)}     &\multicolumn{2}{c}{(pm/V)}       \\
		\hline
		InSCl  &3.848 &3.539 &Indirect&38.13&24.20&22.78&1.34&0.05&7.69 &-0.07 &0.44&-754.41 &31.09 &-54.13 &0.50   \\
		InSBr  &3.899 &3.423 &Indirect&41.10&21.79&29.54&4.02&0.27&8.06 &-0.33 &0.78&-719.20 &29.73 &-37.26 &0.47   \\
		InSI   &3.997 &2.397 &Indirect&43.99&20.47&34.47&7.18&0.56&8.13 &-0.69 &0.64&-764.39 &22.14 &-32.50 &0.34   \\
		InSeBr &4.008 &2.615 &Indirect&30.39&23.51&12.20&0.10&0.04&7.75 &-0.05 &0.64&-1068.12 &29.48&-155.27 &0.55   \\
		InSeI  &4.099 &2.342 &Indirect&34.98&20.53&22.93&3.93&0.30&8.50 &-0.35 &0.47&-1005.14 &22.48&-69.56 &0.41   \\
		InTeCl &4.428 &2.444&Direct  &24.56&6.73 &22.71&-26.7&-2.21&8.38 &2.63 &0.54&135.30&20.24&7.59 &0.65    \\
		InTeBr &4.429 &2.341&Direct  &25.02&6.90 &23.12&-20.6&-1.61&8.75 &1.84 &0.53&117.77&20.38&6.50 &0.64    \\
		InTeI  &4.452 &1.765&Direct  &26.54&7.49 &24.42&-13.6&-0.91&9.13 &1.00 &0.54&115.67&18.80&6.07 &0.55   \\
		\hline\hline
	\end{tabular*}
\end{table*}

To demonstrate the energetic favorability of proposed hexagonal phase for In$XY$ MLs, we performed a comprehensive swarm-intelligence-based structural search using the CALYPSO code \cite{Wang2012,Wang2010}. Computational details are provided in Part 3 of the SM~\cite{supplementary}. Our structural predictions reveal that the hexagonal phases I and II are indeed possible 2D structures for In$XY$ MLs (Fig. S2 and S3~\cite{supplementary}), and phase I serves as the ground-state structure for InSCl, InSBr, and InSeBr MLs. 
Previous experiments had successfully synthesized the metastable phases of 2D transition metal dichalcogenides~\cite{Liu2024a,Liu2018,Jiang2016} with energies 0.04--0.18 eV/atom above their ground states~\cite{Liu2024a,Duerloo2014}. 
Therefore, the hexagonal phase of In$XY$ MLs within an appropriate energy window of 0.05 eV/atom above the ground states should be experimentally accessible.
However, for InO$Y$ ($Y$ = Cl, Br, I) MLs, the hexagonal phase has an energy 0.115--0.271 eV/atom higher than their ground states (Fig. S4(c)~\cite{supplementary}), making their experimental realization difficult.	
We therefore exclude InO$Y$ ($Y$ = Cl, Br, I) MLs from further analysis. 
For the remaining nine In$XY$ ($X$ = S, Se, Te; $Y$ = Cl, Br, I) MLs, the hexagonal phase exists either as the ground state or as a low-energy metastable state, suggesting that synthetic realization is feasible.

Table S2~\cite{supplementary} lists the lattice constants and bond lengths of In$XY$ ($X$ = S, Se, Te; $Y$ = Cl, Br, I) MLs in their lowest-energy states within a hexagonal lattice. In$XY$ MLs have the space group of P3m1 (No. 156) and the $C_{3v}$ point group symmetry. The In ($X$) atom forms bonds with three neighboring $X$ (In) atoms in phases I and II. The lattice constant ($a$) and In–$X$ bond length ($d_{{\rm In}\mbox{-}X}$) grow with increasing radius of the $X$ and $Y$ atoms. In phases I and II, the halogen atom $Y$ forms bonds with three and one adjacent In atoms, respectively. Phase II exhibits shorter In-$Y$ bond lengths ($d_{{\rm In}\mbox{-}Y}$) compared to phase I, indicating stronger In-$Y$ ionic bonding. For example, the $d_{\rm In\mbox{-}I} = 2.680$ \AA\ of InTeI ML in phase II is shorter than $d_{\rm In\mbox{-}I} = 3.104$ \AA\ of InSI ML in phase I, although the Te atom has a larger atomic radius than the S atom.

We examined the structural stability of In$XY$ ($X$ = S, Se, Te; $Y$ = Cl, Br, I) MLs in their lowest-energy states within a hexagonal lattice. 
Table S2~\cite{supplementary} shows that all In$XY$ MLs have a negative formation energy, indicating that their synthesis is energetically favorable. The phonon spectrum (see Fig. S5 (d)~\cite{supplementary}) shows that InSeCl ML is unstable due to the negative frequency at the K point. Other eight In$XY$ MLs exhibit dynamic stability with positive phonon frequencies across the Brillouin zone (see Fig. S5~\cite{supplementary}).   
During the MD simulations at 300 K, all In$XY$ MLs exhibit minor lattice distortion, demonstrating their thermodynamic stability (see Fig. S6~\cite{supplementary}). 
Meanwhile, MD simulations show no spontaneous phase transition of the hexagonal In$XY$ MLs over extended trajectories, demonstrating their kinetic stability under ambient conditions.
Furthermore, the elastic constants of these In$XY$ MLs (Table~\ref{table2}) satisfy the Born criteria for hexagonal lattice: $C_{11}$ $>$ 0, $C_{11}$ $>$ $|C_{12}|$~\cite{mazdziarz2019comment}, exhibiting their mechanical stability.
Thus, after excluding InSeCl ML, the remaining eight In$XY$ MLs were proved to have good structural stability.

In$XY$ ($X$ = S, Se, Te; $Y$ = Cl, Br, I) MLs have excellent flexibility, which is beneficial for their piezoelectricity. 
The $C_{11}$ and $C_{12}$ of In$XY$ MLs show a decreasing trend as the atomic radius of X atom increases.
In the hexagonal lattice, the Young's Modulus ($E$) can be calculated by elastic constants: $E=(C_{11}^2-C_{12}^2)/C_{11}$~\cite{Cadelano2010}.
Table~\ref{table2} shows that InSeBr ML has the lowest $E$ of 12.20 N/m, while other In$XY$ MLs have a $E$ of 22.71--34.47 N/m. 
The $E$ of In$XY$ MLs is much smaller than that of other 2D materials such as graphene (345 N/m)~\cite{Kudin2001}, \ce{MoS_2} (118 N/m)~\cite{Cooper2013}, and h-BN (271 N/m)~\cite{Kudin2001}.  

\begin{figure*}[thb!]
	\centerline{\includegraphics[width=1.0\textwidth]{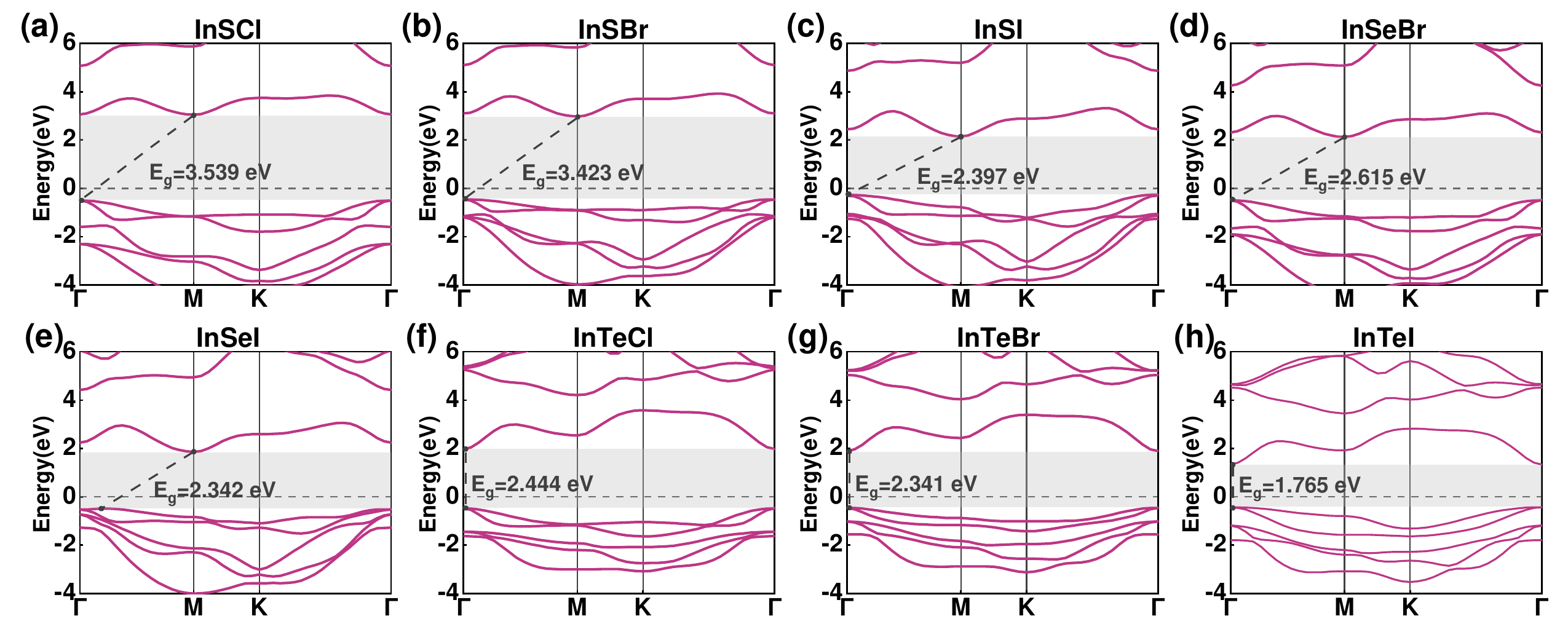}}
	\caption{The band structures of In$XY$ ($X$ = S, Se, Te; $Y$ = Cl, Br, I) MLs.}
	\label{wh3}
\end{figure*}

\subsection{Electronic Structures}
A photocatalyst requires suitable band edge positions straddling the redox potentials of water, which is a premise for water splitting. The standard reduction potential of hydrogen evolution reaction (HER) is $E^{\ce{H+/H2}} = -4.44 + {\rm pH} \times 0.059$, while the oxidation potential of oxygen evolution reaction (OER) is $E^{\ce{O2/H2O}} = -1.23 + E^{\ce{H+/H2}}$. The pH values of 0, 7, and 14 represent the acidic, neutral, and alkaline conditions, respectively. Photocatalysts should have the conduction band minimum (CBM) higher than the reduction potential $E^{\ce{H^{+}/H_{2}}}$ and the valence band maximum (VBM) lower than the oxidation potential $E^{\ce{O_{2}/H_{2}O}}$.

Fig.~\ref{wh3} displays the band structures of eight stable In$XY$ ($X$ = S, Se, Te; $Y$ = Cl, Br, I) MLs.
InTeCl, InTeBr, and InTeI MLs have direct band gaps of 1.765-2.444 eV, with the VBM and CBM being located at the $\Gamma$ point of the Brillouin zone (see Figs.~\ref{wh3} (f-h)). Other In$XY$ MLs have indirect band gaps between 2.342 and 3.539 eV. Their CBM moves to the M point, while their VBM remains at or near the $\Gamma$ point. The band gap of In$XY$ MLs narrows as the atomic masses of the chalcogen atom $X$ or halogen atom $Y$ increase. We analyzed the contribution of different atomic orbitals to electronic states in the band structures. The CBM is mainly contributed by the In-$s$ orbital, while the VBM is formed by $p_x+p_y$ orbitals of chalcogen or halogen atoms (see Fig. S7~\cite{supplementary}). Crystal orbital Hamilton population (COHP)~\cite{Deringer2011} analysis shows that the CBM states are anti-bonding, whereas the VBM states exhibit a mix of bonding and anti-bonding character. We have checked the effect of spin-orbital coupling (SOC) on the electronic structures. In In$XY$ MLs, the SOC effect lifts the degeneracy of the VBM at the $\Gamma$ point, leading to changes in the valence bands, whereas the CBM is only slightly affected. SOC effect will not change the main conclusion in this work, so we did not consider it due to the expensive cost of HSE+SOC calculations.       

\begin{figure}[tbp!]
	\centerline{\includegraphics[width=0.47\textwidth]{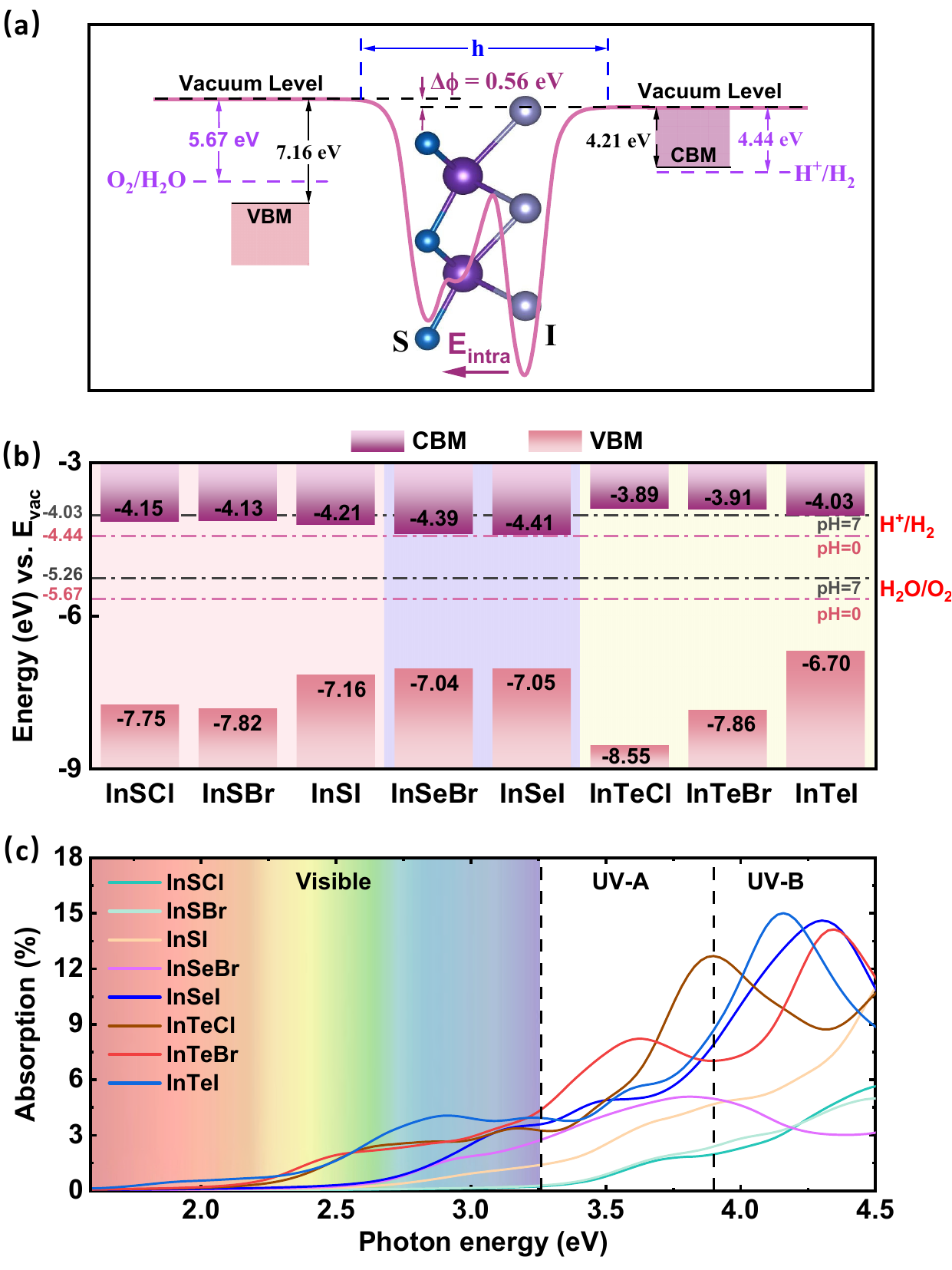}}
	\caption{(a) The average electrostatic potential of InSI ML along the $z$-axis is marked by a pink and solid line. $\Delta\phi$ represents the potential difference between the vacuum level of the S and I atom sides. The direction of intralayer polarization field ($E_{\rm intra}$) is displayed. The $h$ represents the spatial extent of $E_{\rm intra}$ along the $z$-axis. The redox potentials of \ce{H+/H2} and \ce{H2O/O2} at pH = 0 are indicated by purple lines. (b) Band alignments with respect to the vacuum level $E_{\rm vac}$. The redox potentials at pH = 0 (red dashed lines) and pH = 7 (black dashed lines) are shown for comparison. (c) The normalized optical absorbance $A(\omega)$ of In$XY$ ($X$ = S, Se, Te; $Y$ = Cl, Br, I) MLs.}
	\label{wh5}
\end{figure}

Next, we analyzed the arrangement relationship between the band edges (CBM and VBM) and the water redox potentials. The band gaps obtained by HSE06 method are accurate enough in comparison to experiments~\cite{Pela2015}. The asymmetric lattice of In$XY$ MLs results in a vertical polarization $P_z$ and an intralayer polarization field $E_{\rm intra}$. 
There is a potential difference $\Delta\phi$ in the vacuum level $E_{\rm vac}$ between the two sides of In$XY$ MLs (see Fig.~\ref{wh5}(a)). The $E_{\rm intra}$ drives photogenerated electrons to accumulate on the side with a lower $E_{\rm vac}$, facilitating the HER, while holes migrate to the side with a higher $E_{\rm vac}$ for the OER. 
We aligned the CBM and VBM with the corresponding $E_{\rm vac}$ according to the location of electrons and holes. In Fig.~\ref{wh5}(b), the VBM energies of all eight In$XY$ MLs are lower than $E^{\ce{O_{2}/H_{2}O}}$, while their CBM energies are higher than the water reduction potential $E^{\ce{H^{+}/H_{2}}}$ at pH = 0. However, as the pH increases to 7, only InTe$Y$ ($Y$ = Cl, Br, I) MLs have CBM straddling the $E^{\ce{H^{+}/H_{2}}}$.
Therefore, we identified that all eight In$XY$ MLs possess suitable band edge positions for overall water splitting in acidic environments, prompting further investigation into their other photocatalytic properties. 

The hexagonal phase of In$XY$ MLs demonstrates band edge positions suitable for photocatalytic water splitting, along with an intrinsic out-of-plane polarization that facilitates charge separation. In contrast, as detailed in Part 3 of the SM~\cite{supplementary}, the non-hexagonal ground-state structures of In$XY$ MLs exhibit either unsuitable band edge positions or a lack of out-of-plane polarization, which significantly restricts their photocatalytic functionality and are thus excluded from further discussion in this work.

\subsection{Optical absorption}
Photocatalysts harvest sunlight to generate electron-hole pairs for water splitting. Strong optical absorption is an essential condition for photocatalysts. 
Fig.~\ref{wh5}(c) shows the normalized optical absorbance ($A$)~\cite{q36} of In$XY$ ($X$ = S, Se, Te; $Y$ = Cl, Br, I) MLs as a function of the photon energy ($\hbar \omega$), which was obtained using the HSE06 method. 
In the ultraviolet (UV) region, InSeI ML and InTe$Y$ ($Y$ = Cl, Br, I) MLs have higher $A(\omega)$ than other In$XY$ MLs. In the visible zone (1.59--3.26 eV), InTeI ML has the strongest $A(\omega)$ of 4.07\% at 2.92 eV. 
To describe the ability of utilize visible light, we defined the average optical absorption ability $\overline{A}$ in the visible region as 
\begin{equation}
	 \overline{A} = \frac{\int_{1.59}^{3.26} A(\hbar \omega) d(\hbar \omega)}{\int_{1.59}^{3.26} d(\hbar \omega)}
	\label{averA}
\end{equation}
The $\overline{A}$ of InSCl, InSBr, InSI, InSeBr, InSeI, InTeCl, InTeBr and InTeI MLs were estimated as 0.069\%, 0.082\%, 0.35\%, 0.68\%, 0.84\%, 1.43\%, 1.52\%, 1.88\%, respectively. 
Projected band structures in Fig. S7~\cite{supplementary} indicates that photons can excite electrons from the $p$ states in the valence bands to the $s$ states in the conduction bands
according to the selection rules for electric dipole transitions. Fig.~\ref{wh3} reveals that InTe$Y$ ($Y$ = Cl, Br, I) MLs exhibit moderate direct band gaps and flattened valence bands (or high density of states). InTe$Y$ ($Y$ = Cl, Br, I) MLs allow more $p$ states in the valence bands to absorb photons, resulting in stronger light absorption than other In$XY$ MLs. In contrast, InSCl and InSBr MLs exhibit low $A(\omega)$ in the visible region due to their wide indirect band gaps, which result in inefficient utilization of sunlight for water splitting.

\begin{figure}[tbp!]
	\centerline{\includegraphics[width=0.5\textwidth]{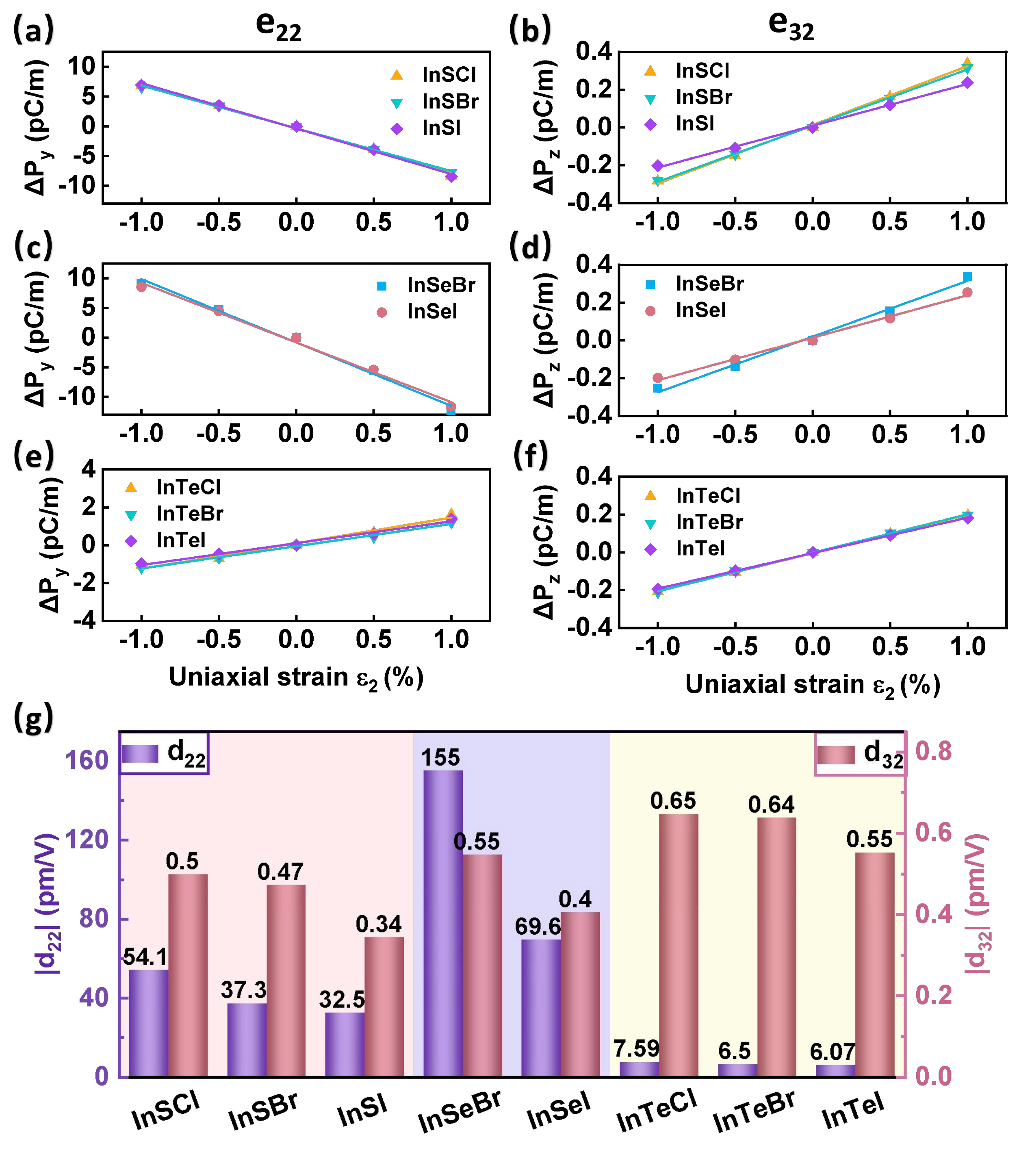}}
	\caption{The change of (a, c, e) in-plane polarization $P_y$ and (b, d, f) out-of-plane polarization $P_z$ of In$XY$ ($X$ = S, Se, Te; $Y$ = Cl, Br ,I) MLs under uniaxial strains $\varepsilon_2$ along the $y$-direction. (g) The absolute value of piezoelectric coefficients $d_{22}$ and $d_{32}$ of In$XY$ MLs.}
	\label{wh4}
\end{figure}

\subsection{Electron-hole separation}
The separation of photoexcited electron-hole pairs is another crucial aspect of photocatalytic water splitting. Both the chalcogen atom $X$ and the halogen atom $Y$ have electronegativity larger than that of the In atom. We performed the Bader charge analysis~\cite{Tang2009} to estimate the transfer charge $\Delta e$ between the In and $X$($Y$) atoms. The $\Delta e$ and vertical coordinate difference $\Delta z$ between the In and $X$ ($Y$) atoms are shown in Table S3~\cite{supplementary}. Two oppositely oriented dipole moments, $D_{\rm X\mbox{-}In}$ and $D_{\rm Y\mbox{-}In}$, are directed from the $X$ and $Y$ atomic layers toward the In atomic layer. Table~\ref{table1} and Fig. S8~\cite{supplementary} shows that the 2D net polarization $P_z$ of In$XY$ ($X$ = S, Se, Te; $Y$ = Cl, Br, I) MLs can be parallel or antiparallel to the $z$-axis due to the competition between $D_{\rm X\mbox{-}In}$ and $D_{\rm Y\mbox{-}In}$. When elements In and $X$ are given but element $Y$ can be changed, the larger the electronegativity difference between element $X$ and $Y$ is, the larger the $P_z$ is. 
We further found the relationship between the polarization $P_z$ of In$XY$ MLs and the elemental properties of the constituent atoms $X$ and $Y$. As shown in Fig. S9~\cite{supplementary}, the $P_z$ exhibits a linear dependence on the atomic structural factor $\bigtriangleup Z \times (r_X+r_Y)$, where $\bigtriangleup Z = Z_X-Z_Y$ represents the difference in atomic numbers between $X$ and $Y$ elements, and ($r_X + r_Y$) denotes the sum of their atomic radii.	

The $P_z$ leads to an intralayer polarization field $E_{\rm intra}$ and the vacuum level difference $\Delta\phi$ between the $X$ and $Y$ atom sides (see Fig.~\ref{wh5}(a) and Fig. S8~\cite{supplementary}). We estimated $E_{\rm intra}$ as $E_{\rm intra} = \Delta\phi /(eh)$ where $h$ is the spatial extent of the intralayer polarization field along the $z$-axis (see Fig.~\ref{wh5}(a)). Table~\ref{table2} shows that both $\Delta\phi$ and $E_{\rm intra}$ are proportional to $P_z$. 

To explore electron-hole separation, we calculated the exciton binding energy $E_b$ of In$XY$ MLs (see Fig. S10 and Table S4~\cite{supplementary}). The $E_b$ of In$XY$ MLs ranges from $0.44$ to $0.78$ eV (see Table~\ref{table2}), which is comparable to that of other 2D photocatalysts such as InSe ML (0.58 eV)~\cite{Wang2018}, CuI ML (0.15 eV)~\cite{Mohebpour2025}, GaSe ML (0.54 eV)\cite{Yang2022}, and WSSe ML (0.83 eV)~\cite{Ju2020}. Moreover, In$XY$ MLs with a high $P_z$ or $E_{\rm intra}$ trend to have a smaller exciton binding energy $E_b$. Thus, a large $P_z$ facilitates the separation of photogenerated electron-hole pairs in In$XY$ MLs.

In$XY$ MLs exhibit out-of-plane piezoelectricity, so external stress can increase $P_z$. 
We calculated the piezoelectric coefficients of In$XY$ MLs using a $1 \times \sqrt{3}$ rectangular cell (see Fig.~\ref{wh1}) with the lattice vectors of $\vec{a}'_1 = a(1, 0)$ and $\vec{a}'_2 = a(0, \sqrt{3})$.
In$XY$ MLs have the $C_{3v}$ point group symmetry and the mirror-symmetry about the $xz$ plane. The piezoelectric tensors can be expressed as
\begin{equation}
e_{ij}=\begin{pmatrix}
	0 & 0 & -e_{22} \\
	-e_{22} & e_{22} & 0 \\
	e_{32} & e_{32} & 0
\end{pmatrix},
\end{equation}
\begin{equation}
	d_{ij}=\begin{pmatrix}
		0 & 0 & -2d_{22} \\
		-d_{22} & d_{22} & 0 \\
		d_{32} & d_{32} & 0
	\end{pmatrix}.
\end{equation}
$e_{22}$ and $e_{32}$ show how in-plane polarization $P_y$ and out-of-plane polarization $P_z$ vary with in-plane  uniaxial strain $\varepsilon_{2}$, respectively. $e_{22}$ and $e_{32}$ were obtained from the slopes of the linear relationship between polarization change and uniaxial strain (see Figs.~\ref{wh4}(a-f)). 
Based on Eq.~\ref{strain}, the $d_{22}$ and $d_{32}$ can be obtained as
\begin{equation}
	d_{22}= \frac{e_{22}} {C_{22}-C_{12}}, \qquad
	d_{32}= \frac{e_{32}} {C_{22}+C_{12}}.
	\label{e3.31}
\end{equation}
For a hexagonal lattice, elastic constant $C_{22}$ is equal to $C_{11}$.
Table~\ref{table2} summarizes the results of $e_{22}$, $e_{32}$, $d_{22}$ and $d_{32}$. When elements In and $X$ are given but element $Y$ can be changed, $e_{32}$ descends as 
the electronegativity difference between the elements $X$ and $Y$ decreases. The similar phenomenon has been founded in the Mo$T$O ($T$ = S, Se, and Te) MLs~\cite{Li2022}. Eq.~\ref{e3.31} shows that $C_{22} + C_{12}$ contributes significantly to the vertical piezoelectric response $d_{32}$. With a small elastic constant $C_{12}$, InTe$Y$ ($Y$ = Cl, Br, I) MLs have greater $d_{32}$ than other In$XY$ MLs (see Fig.~\ref{wh4}(g)). The $d_{32}$ (0.34--0.65 pm/V) of In$XY$ MLs are compared to that of 2D Janus group-III chalcogenide (0.07--0.46 pm/V)~\cite{guo2017enhanced}, but larger than 2D Janus metal dichalcogenides (0.007--0.03 pm/V)~\cite{dong2017large}. Therefore, in-plane stresses can further enhance the vertical polarization $P_z$ of In$XY$ MLs to promote the separation of photogenerated electron-hole pairs. 

In terms of piezoelectric applications, In$XY$ MLs exhibit exceptional in-plane piezoelectricity, with larger $e_{22}$ and $d_{22}$ than those of conventional 2D materials, as shown in Fig.~\ref{wh4}(g).
InSeBr and InSeI MLs have larger in-plane piezoelectric coefficients $e_{22}$ and $d_{22}$ than other In$XY$ MLs, owing to their excellent flexibility (low Young's modulus) and the small difference between the elastic constants $C_{11}$ (or $C_{22}$) and $C_{12}$. The absolute values of $d_{22}$ (6.07--155 pm/V) of In$XY$ MLs are superior to those of ZnO (14.3--26.7 pm/V)~\cite{q21}, 2D metal dichalcogenides (1.50--4.33 pm/V)~\cite{q22}, 2D group-III chalcogenide (GaS, GaSe, GaTe, InS, InSe, InTe) and its Janus structure (1.12--8.47 pm/V)~\cite{guo2017enhanced}. 

\subsection{Electron and hole mobility}
After being separated, photogenerated electrons and holes have to migrate to chemically active sites to participate in the HER and OER. Consequently, high carrier mobility within the photocatalyst is essential for efficient charge transfer. 
During the application of Eq.~\ref{mobility} to estimate carrier mobility, we need to consider the multivalley band structures of InS$Y$ and InSe$Y$ ($Y$ = Cl, Br, I) MLs. In the primitive cell, their CBMs are at the M point of Brillouin zone, which consists of six degenerate valleys. All valleys can be obtained by rotating a reference valley by angles $\theta_i \in [\pi/3, 2\pi/3, \pi, 4\pi/3, 5\pi/3]$. The slope of conduction bands along the M-K and M-$\Gamma$ path is different (see Fig.~\ref{wh3}), indicating that each individual valley has anisotropic electron effective masses. Nevertheless, according to the general derivation given by Sohier et al.~\cite{Sohier2019}, the overall electron mobility of InS$Y$ and InSe$Y$ MLs is still isotropic due to the hexagonal symmetry. The further derivation of Pizzi et al.~\cite{Pizzi2016} indicated that the overall electron mobility $\mu_e$ of InS$Y$ and InSe$Y$ MLs, which is assumed to be averaged over all valleys, is equivalent to averaging the mobility components of the single valley~\cite{Pizzi2016}.

We constructed a $1 \times \sqrt{3}$ rectangular cell to simulate the carrier mobility (see Fig.~\ref{wh1}). The $x$- and $y$-axes are along the zigzag and armchair directions, respectively. When constructing the $1 \times \sqrt{3}$ rectangular cell from the primitive cell, four of the six M points fold into the S(0.5, 0.5, 0) points, while the remaining two fold into the $\Gamma$ point (see Fig. S11~\cite{supplementary}). We calculated the electron mobility components, $\mu_{e, x}^{\Gamma}$ and $\mu_{e, y}^{\Gamma}$, of the single valley at the $\Gamma$ point, along the $x$- and $y$-axes, respectively. The overall electron mobility $\mu_e$ of InS$Y$ and InSe$Y$ MLs can be estimated by ~\cite{Pizzi2016}:
\begin{equation}
	\mu_e = \frac{\mu_{e, x}^{\Gamma} + \mu_{e, y}^\Gamma}{2}.
\end{equation}
On the other hand, both the VBM and CBM of InTe$Y$ ($Y$ = Cl, Br, I) MLs exhibit single valleys at the $\Gamma$ point, which allows direct application of Eq.~\ref{mobility} to estimate their carrier mobility.

\begin{table}[tbh!]
	\caption{\label{table3} The effective mass ($m^{*}$), deformation potential constant ($E_{l}$), and carrier mobility ($\mu^\Gamma_{x}$, $\mu^\Gamma_{y}$) of electron ($e$) and hole ($h$) of the valley at the $\Gamma$ point along the $x$- and $y$-directions.}
	\centering 
	\renewcommand\arraystretch{1.3}
	\begin{ruledtabular}
	\begin{tabular}{cccccccccccc}
		&Carrier  &$m_{x}^{*}$  &$m_{y}^{*}$  &$E_{l,x}$  &$E_{l,y}$  &$\mu^\Gamma_{x}$   &$\mu^\Gamma_{y}$  \\\cline{3-4}\cline{5-6} \cline{7-8}
		
		&         &\multicolumn{2}{c}{($m_{0}$)} &\multicolumn{2}{c}{(eV)}  &\multicolumn{2}{c}{($\rm cm^{2}$/V/s)}     \\
		\hline 
		InSCl  &e &0.460 &0.865 &-5.574&-3.486&121 &81  \\
		       &h &-1.505&-1.568&-3.633&-3.451&28  &27     \\
		InSBr  &e &0.412 &0.759 &-5.996&-0.818&211 &224   \\
		       &h &-2.374&-2.962&-3.230&-3.073&14  &11      \\
		InSI   &e &0.345 &0.630 &-6.822&0.909&250 &269  \\
		       &h &-2.315&-3.241&-3.507&-3.528&12  &8.5      \\
		InSeBr &e &0.378 &0.710 &-2.674&0.458&896 &901  \\
	           &h &-1.363&-1.453&-6.696&-6.435&8   &7     \\
		InSeI  &e &0.338 &0.900 &-6.413&0.488&203 &159   \\
		       &h &-0.348&-1.708&-5.403&-0.538&195 &81       \\
		InTeCl &e &0.173 &0.173 &-8.059&-8.062&269 &269    \\
	           &h &-0.923&-0.959&-3.735&-3.720&43  &42        \\
		InTeBr &e &0.175 &0.176 &-7.922&-7.914&277 &275  \\
		       &h &-1.225&-1.289&-3.648&-3.589&26  &25        \\
		InTeI  &e &0.177 &0.177 &-7.430&-7.732&317 &311   \\
		       &h &-2.537&-2.899&-3.490&-3.361&7   &6   \\
		\end{tabular}
\end{ruledtabular}
\end{table}

\begin{figure}[tbp!]
	\centerline{\includegraphics[width=0.48\textwidth]{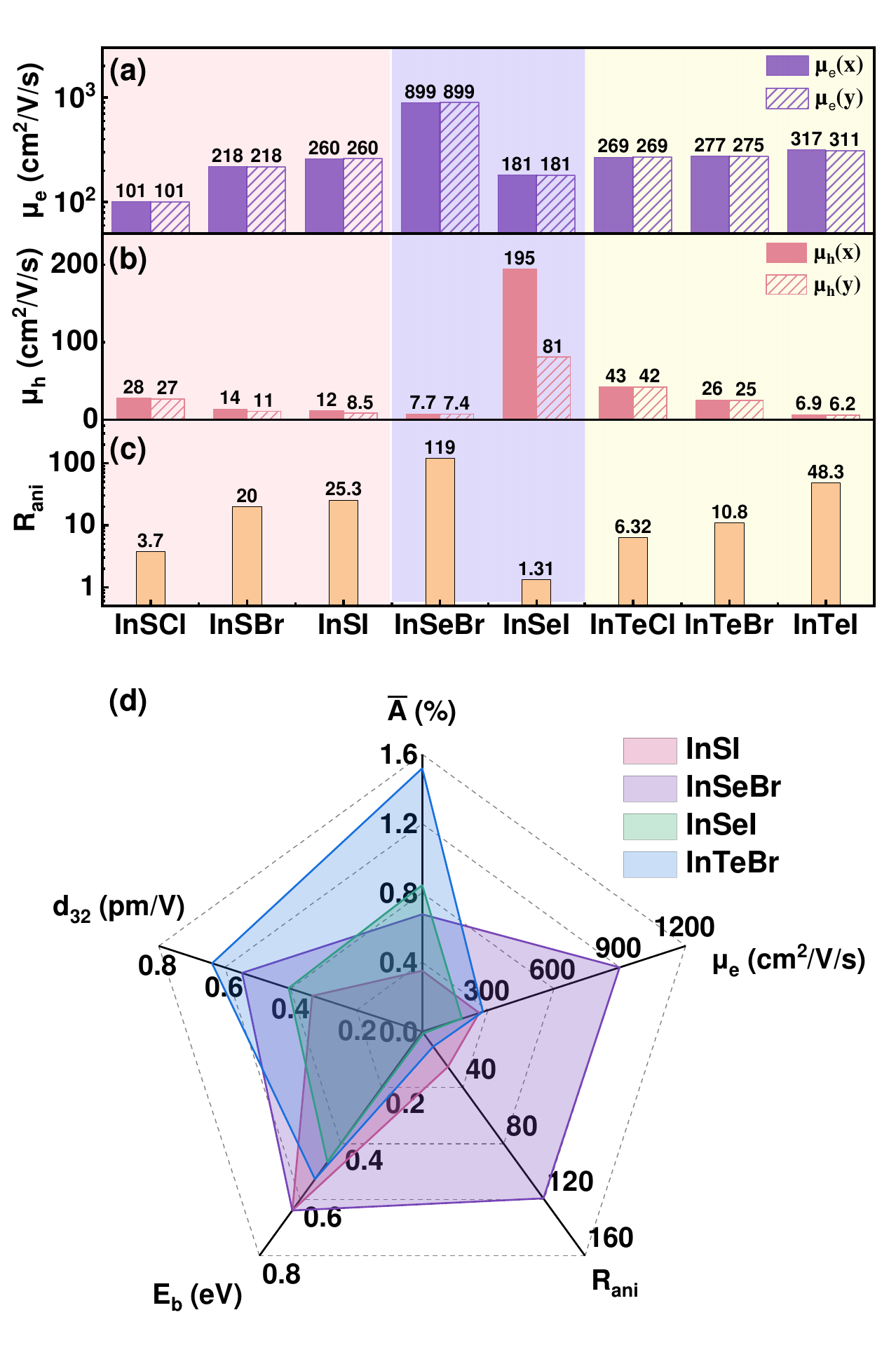}}
	\caption{(a, b) The comparison of electron and hole mobility along the $x$- and $y$- directions in In$XY$ ($X$ = S, Se, Te; $Y$ = Cl, Br, I) MLs. (c) The carrier mobility anisotropy ratio $R_{ani}$ of In$XY$ MLs. (d) The radar map illustrates five key factors highly correlated with piezo-photocatalyst performance.}
	\label{wh6}
\end{figure}

Table~\ref{table2} and \ref{table3} list the elastic constants, the carrier effective mass, and the deformation potential (DP) constants. InTe$Y$ ($Y$ = Cl, Br, I) MLs possess nearly identical electron effective masses and DP constants along the $x$- and $y$-directions. Combined with the previous analysis of InS$Y$ and InSe$Y$ MLs, we confirmed that all In$XY$ ($X$ = S, Se, Te; $Y$ = Cl, Br, I) MLs have in-plane isotropic electron mobility ($\mu_e$), as illustrated in Fig.~\ref{wh6}(a). The high $\mu_e$ of InSeBr ML (899 cm$^{2}$/V/s) is mainly attributed to its small DP constants. The electron mobility of In$XY$ MLs is close to or higher than that of other 2D photocatalysts predicted with the Eq.~\ref{mobility}, such as \ce{TlAsS2} (209--600 cm$^{2}$/V/s)~\cite{Fang2025}, In$_2X_3$ ($X$ = S, Se) (340--487 cm$^{2}$/V/s)~\cite{Rivera2025}, WSSe$_2$ (125 cm$^{2}$/V/s)~\cite{Ju2020}, and ZrO$X_2$ ($X$ = Br, I) (80--132 cm$^{2}$/V/s)~\cite{Yang2024}.

In comparison, Fig.~\ref{wh6}(b) shows that the hole transport properties of In$XY$ MLs are inferior to the electron one. The overall hole effective mass is larger than that of the electron (see Table~\ref{table3}). Except for InSeI ML, In$XY$ MLs have isotropic hole mobility. The VBM of InSeI ML is located near the $\Gamma$ point (see Fig.~\ref{wh3}(e)), and its hole effective mass is highly anisotropic. 

We defined the anisotropy ratio as $R_{ani}= |\overline{\mu}_e/ \overline{\mu}_h|$, which is to measure the difference in electron and hole transport, where $\overline{\mu}_{e}$ and $\overline{\mu}_{h}$ are the averaged electron and hole mobility, respectively.
Except for InSeI ML, In$XY$ MLs have a substantial $R_{ani}$, as seen in Fig.~\ref{wh6}(c). For photocatalytic water splitting, high $R_{ani}$ is beneficial to decreasing the recombination for photoexcited carriers~\cite{meng2018two,li2019enhanced,sun2019flexible}.

Based on previous results, we systematically evaluated the potential of In$XY$ MLs ($X$ = S, Se, Te; $Y$ = Cl, Br, I) as piezo-photocatalysts by analyzing key performance-related factors: average visible-light absorption ability $\overline{A}$, out-of-plane piezoelectric coefficient $d_{32}$, exciton binding energy $E_b$, electron mobility $\mu_e$, and carrier mobility anisotropy ratio $R_{ani}$. Notably, InSCl and InSBr MLs exhibit poor visible-light absorption.
InTe$Y$ ($Y$ = Cl, Br, I) MLs exhibit similar properties. 
Therefore, we focus our comparative analysis on four representative systems: InSI, InSeBr, InSeI, and InTeBr MLs, as summarized in Fig.~\ref{wh6}(d).

Our analysis reveals that, while individual In$XY$ MLs excel in specific aspects, they often exhibit compensating drawbacks. For instance, InTeBr ML shows strong visible-light absorption (high $\overline{A}$) and a large out-of-plane piezoelectric coefficient ($d_{32}$), but a low carrier mobility anisotropy ($R_{ani}$). InSeBr ML outperforms others in $d_{32}$, electron mobility ($\mu_e$), and $R_{ani}$, but its visible-light absorption ($\overline{A}$) is only moderately competitive.
InSI and InSeI MLs exhibit inferior overall performance compared to InSeBr and InTeBr MLs. To combine the advantages of individual monolayers into a single system, we proposed heterojunction engineering. Among these candidates, InSI, InSeBr, and InSeI MLs exhibit small lattice mismatches (see Table~\ref{table2}), which is favorable to construct vdW heterojunctions.
Among three possible heterojunctions, both InSeBr/InSeI and InSI/InSeI heterojunctions fail to maintain the type-II band alignment and excellent piezoelectricity in their low-energy stacking configurations (see Fig. S16~\cite{supplementary}), which hinders them as efficient piezo-photocatalysts.
Consequently, we focused on the InSI/InSeBr vdW heterojunction to explore its photocatalytic performance.

\begin{table*}[tbh!]
	\caption{\label{table4}
		Calculated properties of InSI/InSeBr heterojunction: lattice constant ($a$), indirect band gap ($E_{g}$) calculated by HSE06 method, relaxed-ion elastic stiffness coefficients ($C_{11}$ and $C_{12}$), the piezoelectric coefficients ($e_{22}$, $e_{32}$, $d_{22}$, $d_{32}$), carrier mobility ($\mu_x$, $\mu_y$) of electrons and holes, carrier mobility anisotropy ration ($R_{ani}$), and tunneling probability ($T_{B}$).}
	\renewcommand\arraystretch{1.3}
	\centering 
	\begin{tabular*}{1\textwidth}{@{\extracolsep{\fill}}*{24}{c}}
		\hline \hline
		&$a$  &$E_{g}$  &$C_{11}$ &$C_{12}$  &$e_{22}$  &$e_{32}$  &$d_{22}$  &$d_{32}$  &$\mu_{e,x}$ &$\mu_{e,y}$  &$\mu_{h,x}$  &$\mu_{h,y}$  &$R_{ani}$  &$T_{B}$     \\\cline{4-5} \cline{6-7} \cline{8-9}
		\cline{10-13}		
		& (\AA)  &(eV) &\multicolumn{2}{c}{(N/m)}  &\multicolumn{2}{c}{(pC/m)}     &\multicolumn{2}{c}{(pm/V)} &\multicolumn{4}{c}{(cm$^{2}$/V/s)} & &(\%)    \\
		\hline 
		I-Se(AA)  &3.977 &2.330 &79.4 &42.4 &1641.3  &-66.5  &44.4&-0.55&437&437&24&15&22.4&4.33 \\
		I-Se(AB$^{\prime}$) &3.978 &2.298 &78.6 &41.5 &-290.6   &-68.4  &-7.8 &-0.57&410&410&22&15&22.2&4.63 \\
		S-Se(AC$^{\prime}$) &3.986 &2.314 &75.6 &40.8 &-1789.0   &-4.3   &-51.5 &-0.04&434&434&17&13&29.0&4.47 \\
		\hline\hline 
	\end{tabular*}
\end{table*}

\subsection{InSI/InSeBr heterojunction}
InSI/InSeBr heterojunction has four kinds of interface structures: S-Br, I-Br, S-Se, and I-Se interfaces. We constructed six possible stacking configurations, namely AA, AA$^{\prime}$, AB, AB$^{\prime}$, AC, and AC$^{\prime}$, for each interface by translating and rotating the bottom InSeBr ML relative to the top InSI ML (see Figs. S12--S15~\cite{supplementary}). The binding energy ($\Delta E_{b}$) of heterojunction is defined as  
\begin{equation}
	\Delta E_{b} = E_{\rm InSI/InSeBr} - E_{\rm InSI} - E_{\rm InSeBr},
\end{equation} 
where $E_{\rm InSI/InSeBr}$, $E_{\rm InSI}$, and $E_{\rm InSeBr}$ are the energies of InSI/InSeBr heterojunction, isolated InSI ML, and isolated InSeBr ML, respectively. We relaxed the lattice structure for all stacking configurations. We have labeled the interlayer distances and the binding energies ($\Delta E_{b}$) for each configuration, as shown in Figs. S12--S15~\cite{supplementary}. 
The S-Se and I-Se interfaces in the InSI/InSeBr heterojunction show lower binding energies $\Delta E_{b}$ than the other two interfacial configurations.

\begin{figure}[tbp!]
	\centerline{\includegraphics[width=0.45\textwidth]{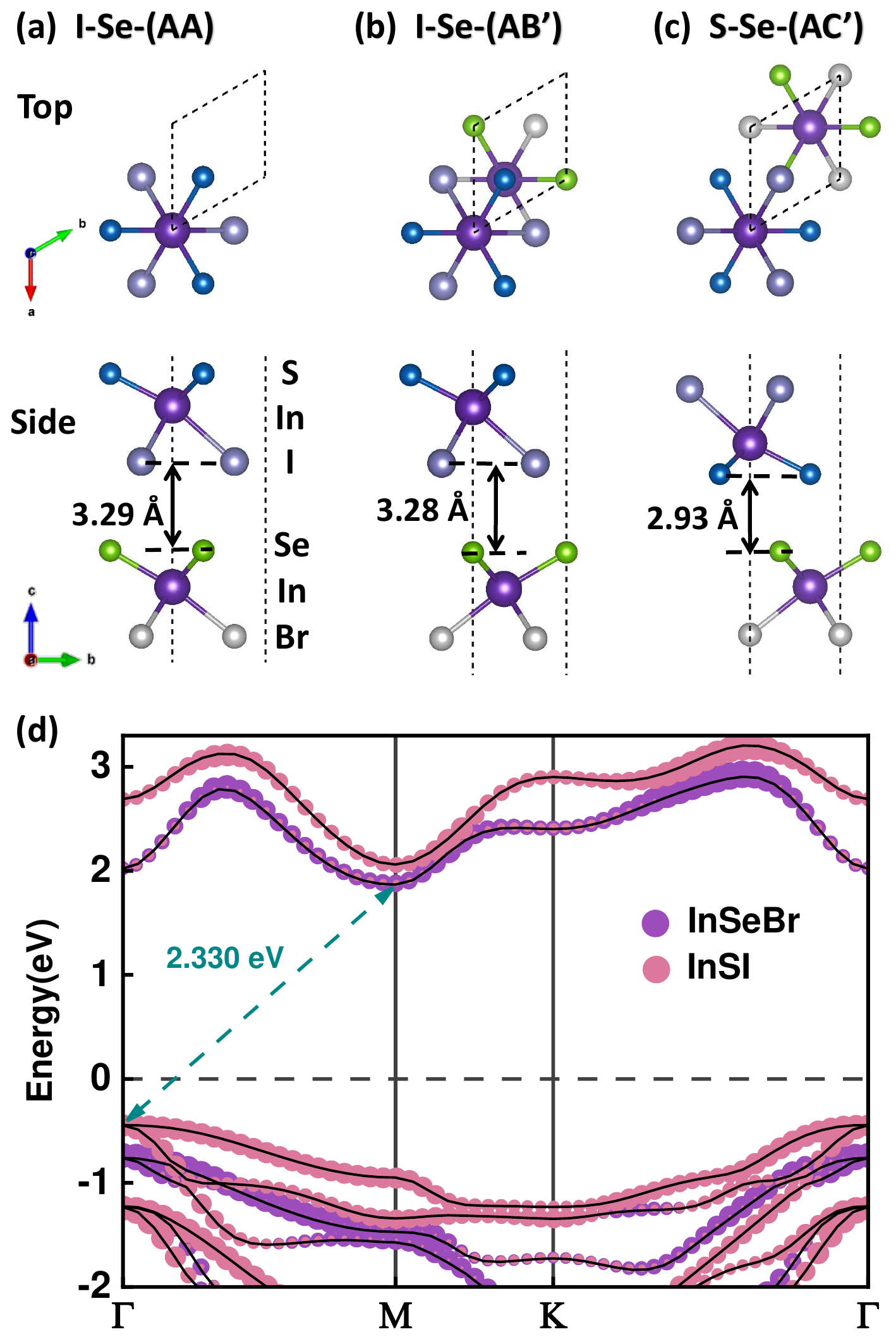}}
	\caption{The top and side view of optimized structures of heterojunction (a) I-Se(AA), (b) I-Se(AB$^{\prime}$), and (c) S-Se(AC$^{\prime}$). (d) The projected band structure by the HSE06 functional of heterojunction I-Se(AA). The pink and purple bubbles represent the contribution of electronic states of the InSI and InSeBr MLs, respectively.}
	\label{wh7}
\end{figure}

The construction of the InSI/InSeBr heterojunction is to combine the advantages of individual monolayers to improve photocatalytic performance. InSI/InSeBr heterojunction should preserve the excellent piezoelectricity of isolated InSeBr and InSI MLs. According to Figs.~\ref{wh4} (a-d), if the top and bottom monolayers in the heterojunction are oriented in the same directions, their changes in polarization ($P_y$ and $P_z$) under uniaxial strains will have the same signs, thus achieving an enhanced piezoelectricity. Therefore, we chose the InSI/InSeBr heterojunction with the I-Se interface in the AA stacking configuration (labeled as heterojunction I-Se(AA)), as shown in Fig.~\ref{wh7}(a). The S, In, and I atoms of the top InSI ML are on top of the Se, In, and Br atoms of the bottom InSeBr ML, respectively. Its lattice constant $a = 3.977$ \AA\ is smaller than that of the isolated InSI and InSeBr MLs, due to interlayer coupling. Both InSI and InSeBr MLs have the same orientation along the $y$- and $z$-axes in this configuration. Table~\ref{table4} shows that the heterojunction I-Se(AA) has a absolute value of piezoelectric coefficients $|e_{22}|$ = 1641.3 pC/m and $|e_{32}|$ = 66.5 pC/m, which are higher than its monolayer value. Meanwhile, heterojunction I-Se(AA) exhibits enhanced elastic constants ($C_{11} = 79.4$ N/m and $C_{12} = 42.4$ N/m). According to Eq.~\ref{e3.31}, its $|d_{22}|$ of 44.4 pm/V is reduced compared to its monolayer counterpart, while its $|d_{32}|$ (0.55 pm/V) remains at a similar value.

To justify the above analysis, we rotated the bottom InSeBr ML by 180$^\circ$ around the $z$-axis and then shifted InSeBr ML to find a relatively stable AB$^{\prime}$ configuration (labeled as heterojunction I-Se(AB$^{\prime}$) in Fig.~\ref{wh7}(b)). In this configuration, the top InSI ML and bottom InSeBr ML exhibit oppositely oriented variations in their in-plane polarizations, resulting in a minor net change in the total in-plane polarization under uniaxial strains $\varepsilon_2$. According to Eq.~\ref{eij}, heterojunction I-Se(AB$^{\prime}$) exhibits weak in-plane piezoelectricity with $|e_{22}|$ = 290.6 pC/m, while its vertical piezoelectricity ($|e_{32}|$ = 68.4 pC/m) is comparable to that of heterojunction I-Se(AA). 
If we further rotated the top InSI ML by 180$^\circ$ around the $x$-axis, we could obtain the AC$^{\prime}$ configuration with the S-Se interface (labeled as heterojunction S-Se(AC$^{\prime}$) in Fig.~\ref{wh7}(c)). 
Under uniaxial strain $\varepsilon_2$, the top InSI ML and the bottom InSeBr ML exhibit same oriented variations in their in-plane polarizations, while their out-of-plane polarizations show oppositely oriented variations. As a result, S-Se(AC$^{\prime}$) exhibits a high in-plane piezoelectric coefficient $|e_{22}|$ but low out-of-plane $|e_{32}|$, as shown in Table~\ref{table4}. Consequently, heterojunction I-Se(AA) exhibits superior overall piezoelectric performance, and we will focus on its photocatalytic performance.

Fig.~\ref{wh7}(d) shows that the heterojunction I-Se(AA) has an indirect band gap of 2.330 eV, which is smaller than that of isolated InSI (2.397 eV) and InSeBr (2.615 eV) MLs. This indicates that it has an enhanced light absorption capacity. The electronic structures of InSI ML and InSeBr ML are well preserved in the heterojunction. The CBM and VBM are located at the M and $\Gamma$ points, respectively. Moreover, the CBM and VBM are dominated by the electronic states in InSeBr and InSI MLs, respectively. Thus, heterojunction I-Se(AA) has a type-II band alignment. Considering the expensive cost of the HSE06 method, we calculated the band structure of all stacking configurations using the PBE functionals (see Figs. S12--S15~\cite{supplementary}). The type-II band alignment is a universal feature in the InSI/InSeBr heterojunctions. Meanwhile, given the interface of InSI/InSeBr heterojunction, the type-II band alignment is robust to the translation and rotation between top and bottom  layers, due to weak interlayer coupling. 
   
\subsection{The Z-scheme carrier transfer pathway in InSI/InSeBr heterojunction}
\begin{figure}[tbp!]
	\centerline{\includegraphics[width=0.45\textwidth]{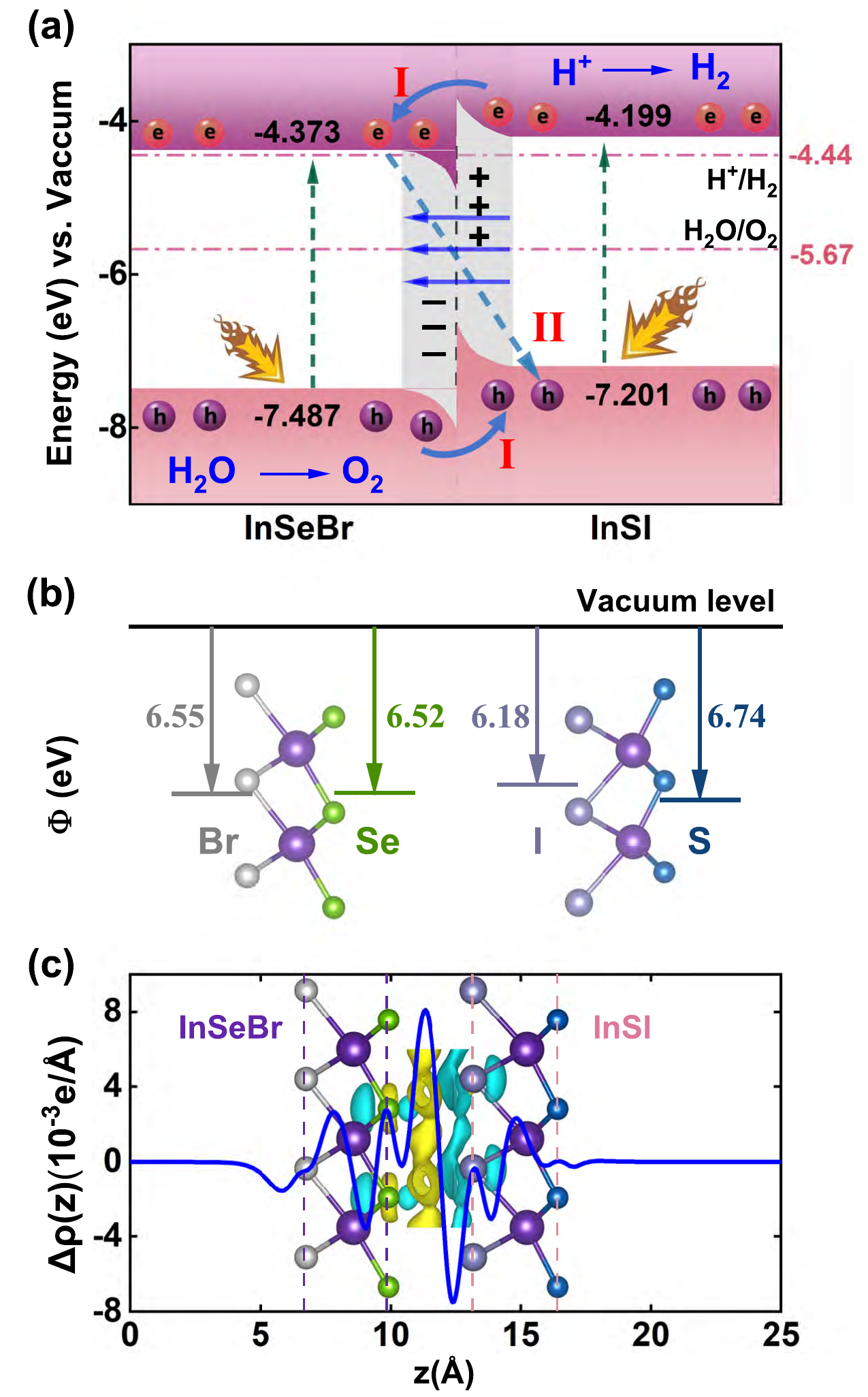}}
	\caption{(a) The edge position diagram of InSI and InSeBr MLs in the heterojunction with I-Se interface. The light gray area is the space charge area. The interfacial electric field is indicated by blue arrows. (b) The work function ($\Phi$) of isolated InSI ML and InSeBr ML before contact. (c) The plane-integrated electron density difference along the $z$-direction for InSI/InSeBr heterojunction. The yellow and green regions represent electron accumulation and depletion, respectively. The isosurface value is 0.00006 e/\AA$^3$. }
	\label{wh8}
\end{figure}

Fig.~\ref{wh8}(a) demonstrates that the band edge positions of the heterojunction I-Se(AA) straddle the redox potentials of water at pH = 0, suggesting that photogenerated electrons and holes can drive the HER and OER, respectively. 

The charge transfer between the InSI and InSeBr MLs generates an interfacial electric field $E_{\rm inter}$ at the I-Se interface. Fig.~\ref{wh8}(b) shows the work function of individual InSI and InSeBr MLs before contact. We estimated the work function as $\Phi = E_{\rm vac} - E_{f}$, where $E_{\rm vac}$ and $E_{f}$ are the vacuum level and the Fermi energy, respectively. The $\Phi$ of the I atom side of the InSI ML is smaller than that of the Se atom side of the InSeBr ML. When InSI and InSeBr MLs come together to form the heterojunction with the I-Se interface, electrons are inclined to transfer from InSI to InSeBr until the Fermi levels are agreed consistent. 

To display the interlayer charge transfer, we adopted the 
planar-average charge density difference with following definition: 
\begin{equation}
	\Delta\rho = \rho_{\rm InSI/InSeBr} - \rho_{\rm InSI} - \rho_{\rm InSeBr},
\end{equation}
where $\rho_{\rm InSI/InSeBr}$, $\rho_{\rm InSI}$, and $\rho_{\rm InSeBr}$ are the plane-averaged electron densities of the InSI/InSeBr heterojunction, the component InSI, and InSeBr MLs, respectively. The charge difference distribution $\Delta\rho$ in Fig.~\ref{wh8}(c) verifies the accumulation and depletion of electrons near the Se atom and I atom sides during the formation of the I-Se interface. In InSeBr ML, the intrinsic dipole moment points from the Se atom to the Br atom, and interlayer charge transfer further enhances this polarization. The blue arrows in Fig.~\ref{wh8}(a) indicate that the interfacial electric field $E_{\rm inter}$ points from InSI ML to InSeBr ML. Meanwhile, $E_{\rm inter}$ causes a downward band bending of InSeBr ML and an upward band bending of InSI ML near the I-Se interface. 

We confirmed that the above method can reliably predict the direction of $E_{\rm inter}$ in InSI/InSeBr heterojunctions with other interfaces (see Figs. S12--S15~\cite{supplementary}). Heterojunctions with a given interface exhibit consistent band alignment, interlayer charge transfer direction, and interfacial electric field ($E_{\rm inter}$) orientation, regardless of interlayer displacement or rotation. For clarity, we specifically illustrated the interfacial charge transfer and the direction of $E_{\rm inter}$ for the AA configuration for each interface.

In the heterojunction I-Se(AA), the $E_{\rm inter}$ accelerates photogenerated carrier transfer through the Z-scheme pathway. Upon photon excitation, electrons in the valence bands of both InSeBr and InSI MLs are excited to their conduction bands, leaving holes in the valence bands. The intralayer polarization field $E_{\rm intra}$ separates the photogenerated electron-hole pairs in each monolayer. 
Charged carriers have two possible transfer pathways: (I) electrons transfer from the conduction band of InSI ML to that of InSeBr ML; holes transfer from the valence band of InSeBr ML to that of InSI ML, and (II) the recombination between electrons in InSeBr's conduction band and holes in InSI's valence band. The interfacial electric field ($E_{\rm inter}$) and band bending selectively suppress pathway I, while promoting pathway II. 
Additionally, the Coulomb repulsion among the conduction bands' electrons (or among the valence bands' holes) also suppress the pathway I.
This results in that strongly oxidative holes assemble in the valence band of InSeBr ML to oxidize the water to produce O$_2$, while strongly reductive electrons assemble in the conduction band of InSI ML to reduce the water to produce H$_2$ (see Fig.~\ref{wh8}(a)). 
Thus, heterojunction I-Se(AA) exhibits a Z-scheme pathway for carrier transfer, which can spatially separate photogenerated electron-hole pairs. 

In the heterojunction I-Se(AA), InSI ML acts as a reduced photocatalyst, while InSeBr ML acts as an oxidized photocatalyst.  
However, photogenerated electrons may reduce InSI ML first rather than water, the holes may oxidize InSeBr ML instead of water. We adopted the method proposed by Chen et al.~\cite{chen2012thermodynamic} to demonstrate that InSeBr ML is resistant to hole oxidation and InSI ML remains stable against electron reduction in water under light (see Part 11 of SM~\cite{supplementary}). Thus, heterojunction I-Se(AA) can resist photocorrosion in an aqueous solution. 

\begin{figure}[tbp!]
	\centerline{\includegraphics[width=0.45\textwidth]{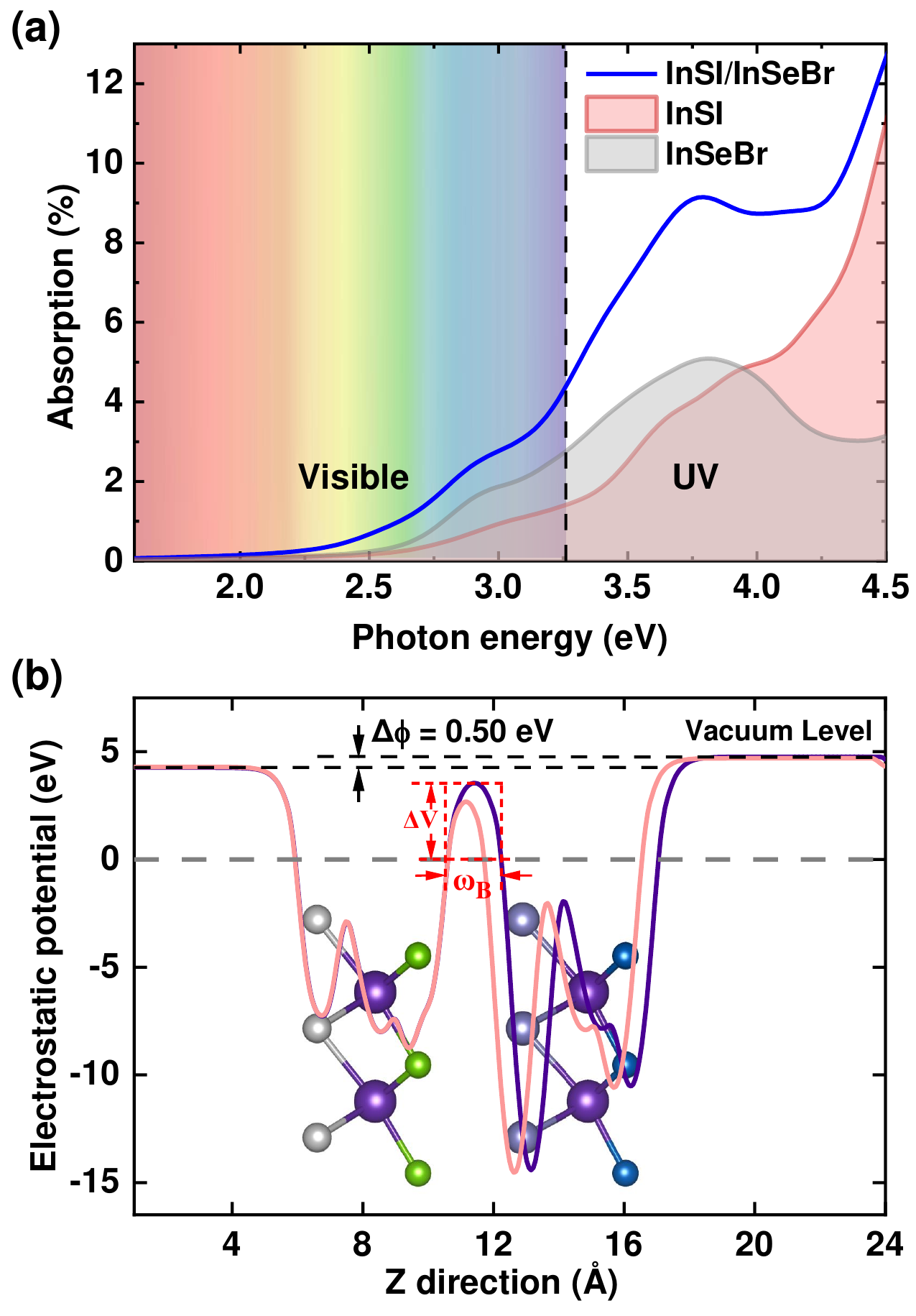}}
	\caption{(a) Absorption spectra comparison between InSI/InSeBr heterojunction and isolated InSI ML and InSeBr ML. (b) Average electrostatic potential in the plane normal to InSI/InSeBr heterojunction. The rectangular box with dashed red line indicates the tunneling barrier. The Purple and pink lines stand for the electrostatic potential with interlayer distance of 3.29 \AA\ and 2.79 \AA, respectively. $\Delta\phi$ represents the vacuum level difference between the two sides of the heterojunction.}
	\label{wh9}
\end{figure}

\subsection{The photocatalytic properties of InSI/InSeBr heterojunction} 
Fig.~\ref{wh8}(a) shows that the CBM energy of InSI ML in the heterojunction rises by 0.011 eV compared to individual InSI ML (see Fig.~\ref{wh5}(b)),  while the VBM energy of InSeBr ML declines by 0.45 eV compared to individual InSeBe ML. Thus, heterojunction I-Se(AA) exhibit stronger redox capacity than its component monolayers.

Concerning the utilization of sunlight, Fig.~\ref{wh9}(a) shows that heterojunction I-Se(AA) exhibits a significantly elevated optical absorption $A(\omega)$ across a broader spectral range, compared to isolated InSI and InSeBr MLs. We analyzed the solar-to-hydrogen (STH) efficiency $\eta_{\rm STH}$ with the approach proposed by Fu et al~\cite{Fu2018}. Part 12 of the SM~\cite{supplementary} provides detailed information. The $\eta_{\rm STH}$ values of isolated InSI ML and InSeBr ML are 9.63\% and 5.22\%, respectively. Heterojunction I-Se(AA) shows a high STH efficiency of 11.12\%, exceeding the 10\% threshold required for commercial applications of photocatalytic water splitting. 

Table~\ref{table4} lists the carrier mobility of heterojunction I-Se(AA). 
Its electron mobility reaches 437 cm$^2$/V/s, falling between that of isolated InSI and InSeBr MLs. The CBM of heterojunction I-Se(AA) is mainly contributed by InSeBr ML. The electron DP constants are larger than those of isolated InSeBr ML (see Table S5~\cite{supplementary}), which is the main reason for reduced electron mobility. Thus, interlayer coupling has a negative effect on the electron mobility of heterojunction I-Se(AA). However, the carrier mobility anisotropy ratio is $R_{ani} = 22.4$, which is high to suppress the combination of electron-hole pairs. Moreover, heterojunction I-Se(AB$^{\prime}$) and S-Se(AC$^{\prime}$) have carrier mobility comparable to that of I-Se(AA), indicating that the interlayer translation and rotation have a limited influence on the carrier transport of InSI/InSeBr heterojunction. 

The overall carrier transport efficiency of heterojunction is determined not only by in-plane carrier mobility but also by interfacial tunneling efficiency. A low interfacial tunneling probability will prevent the photogenerated carrier from crossing the interface of heterojunction, inevitably lowering the photocatalytic efficiency of the Z-scheme heterojunction. The real potential barrier can be simplified to a square barrier with a width ($\omega_{B}$) and a height ($\Delta V$)~\cite{ju2019high,Yu2024}, which can be obtained by electrostatic potential measurement (Fig.~\ref{wh9}(b)). The tunneling probability ($T_{B}$) is evaluated using the formula:
\begin{equation}
	T_{B} = \exp\left\{-2 \frac{\sqrt{2m_0 \Delta V}}{\hbar} \times \omega_{B}\right\},
	\label{e3.33}
\end{equation}
Where $m_0$ and $\hbar$ are the mass of free electrons and reduced Planck's constant. 
Heterojunction I-Se(AA) has $\Delta V$ and $\omega_{B}$ values of 3.54 eV and 1.628 \AA, thereby a the $T_B$ of 4.33\%. The $T_B$ of other configurations is shown in Table~\ref{table4}, which is superior to the heterojunction photocatalyst \ce{g-C3N4 / MoS2} (3.67\%)~\cite{ju2019high}.

Strengthening the interlayer coupling can raise the tunneling probability $T_B$. When the interlayer spacing of heterojunction I-Se(AA) is reduced by 0.5 \AA, $\Delta V$ and $\omega_{B}$ decrease to 2.691 eV and 1.107 \AA, respectively, and $T_B$ increases to 15.56\%, which facilitates the carrier transport across the I-Se interface. Interlayer distance engineering may become an essential strategy for optimizing quantum transport in vdW heterojunctions.

\section{CONCLUSION}
In summary, our first-principles calculations reveal that eight Janus In$XY$ MLs, including InSCl, InSBr, InSI, InSeBr, InSeI, InTeCl, InTeBr, and InTeI MLs, exhibit exceptional potential for photocatalytic water splitting, possessing optimal band gaps (1.765--3.539 eV), suitable band edges straddling the redox potentials, strong vertical polarization (up to 26.7 pC/m), low exciton binding energies (0.44--0.78 eV), excellent electron mobility (101--899 cm$^2$/V/s), and large anisotropy in carrier transport. 
Their out-of-plane polarization $P_z$ can be well described by a structural factor involving the atomic number and atomic radius of $X$ and $Y$ atoms.   
Furthermore, these In$XY$ MLs exhibit remarkable piezoelectric properties with large in-plane ($|e_{22}| = $115.67--1068.12 pC/m, $|d_{22}|= $6.07--155.27 pm/V) and out-of-plane ($|e_{32}| =$ 18.80--31.09 pC/m, $|d_{32}| =$ 0.34--0.65 pm/V) coefficients. Through rational design of 2D Janus vdW heterojunctions using the optimal InSI and InSeBr MLs, we demonstrated that the I-Se(AA) configuration exhibits record-high piezoelectric coefficients ($|e_{22}|$ = 1641.3 pC/m, $|e_{32}|$ = 66.5 pC/m), while the interfacial electric fields enable efficient electron-hole separation through a Z-scheme charge transfer pathway. Moreover, the heterojunction I-Se(AA) exhibits enhanced redox potentials, a STH efficiency of 11.12\%, and broadened visible-light absorption compared to constituent monolayers. Although interlayer coupling reduces the electron mobility of I-Se(AA), this is compensated by enhanced hole mobility, spatial separation of photogenerated carriers with suppressed recombination, and enhanced tunneling probability. These findings reveal a fundamental competition between tunneling probability (positively correlated with coupling strength) and electron mobility (inversely correlated) in InSI/InSeBr heterojunction, suggesting interface engineering as a critical strategy for synergistic optimization in future design of heterojunction.

\begin{acknowledgments}
We acknowledge Dr. Busheng Wang (Yanshan university) and Dr. Sheng Wang (Yanshan University) for their fruitful discussions.
This work was supported by the National Natural Science Foundation of China (No. 11904313), the Scientific Research Foundation of the Higher Education of Hebei Province, China (No. BJ2020015), Hebei Natural Science Foundation (A2022203006), the Science and Technology Project of Hebei Education Department (BJK2022002), and the Innovation Capability Improvement Project of Hebei province (Grant No. 22567605H). The numerical calculations in this paper have been done on the supercomputing system in the High Performance Computing Center of Yanshan University.
\end{acknowledgments}

\nocite{*}


\providecommand{\noopsort}[1]{}\providecommand{\singleletter}[1]{#1}%

\end{document}